%% file: paper.tex
\documentclass[10pt,twocolumn,conference]{IEEEtran}
\IEEEoverridecommandlockouts
\usepackage{cite}
\usepackage{amsmath,amssymb,amsfonts}
\usepackage{algorithmic}
\usepackage{graphicx}
\usepackage{textcomp}
\usepackage[dvipsnames]{xcolor}
\usepackage{hyperref}
\usepackage{svg}
\usepackage[nolist]{acronym} 
\usepackage{balance}
\usepackage{todonotes}
\usepackage{listings}
\usepackage{fancyhdr}

\usepackage{etoolbox}
\makeatletter
\patchcmd{\@makecaption}
  {\scshape}
  {}
  {}
  {}
\makeatletter
\patchcmd{\@makecaption}
  {\\}
  {.\ }
  {}
  {}
\makeatother
\def\tablename{Table}


\input{macros/macrosIEEE}

\input{macros/macros}
\input{acronyms}

\fancypagestyle{IEEEtitlepagestyle}{
    \fancyhf{} 
     
    \fancyhead[C]{
        \small Published at the \emph{1st KuVS Workshop on Resilient Networks and Systems (ReNeSys)}, Sept. 2025, Ilmena/Germany
        }
}

\def\BibTeX{{\rm B\kern-.05em{\sc i\kern-.025em b}\kern-.08em
    T\kern-.1667em\lower.7ex\hbox{E}\kern-.125emX}}
\begin{document}

\title{SENSOR: A Co\underline{s}t-\underline{E}fficient Ope\underline{n}-\underline{S}ource Fl\underline{o}w Monito\underline{r}ing Platform
\thanks{This work was supported by the bwNET2.0 project which is funded by the Ministry of Science, Research and the Arts Baden-Württemberg (MWK). The authors alone are responsible for the content of this paper.}
}

\author{
    \IEEEauthorblockN{
        Gabriel~Paradzik\IEEEauthorrefmark{3}\IEEEauthorrefmark{4},
        Benjamin~Steinert\IEEEauthorrefmark{3}\IEEEauthorrefmark{4}, 
        Heinrich~Abele\IEEEauthorrefmark{4}, 
        Michael~Menth\IEEEauthorrefmark{3}
	}
    \IEEEauthorblockA{\IEEEauthorrefmark{3}
        University~of~Tübingen,
        Chair~of~Communication~Networks,
        Tübingen,
        Germany\\
    }
    \IEEEauthorblockA{\IEEEauthorrefmark{4}
        University~of~Tübingen,
        Zentrum~für~Datenverarbeitung,
        Tübingen,
        Germany
    }
}

\maketitle

\input{chapters/00-abstract.tex}


\input{content}

\balance 

\bibliographystyle{IEEEtranS}
\bibliography{literature.bib}

\end{document}

%% file: macros/macrosIEEE.tex
\newlength{\figsize}
\newlength{\subfigwidth}
\newlength{\subfiglabelwidth}


%% file: macros/macros.tex
\newcommand{\new}[1]{\textcolor{orange}{#1}}


%% file: acronyms.tex
\begin{acronym}
\acro{ie}[IE]{Information Element}
\acro{dpi}[DPI]{deep packet inspection}
\acro{netsa}[NetSA]{Network Situational Awareness}
\acro{DoS}[DoS]{Denial of Service}
\acro{IoC}{indicator of compromise}
\acro{vm}[VM]{virtual machine}
\acro{nic}[NIC]{network interface card}
\acrodefplural{IoC}[IoCs]{indicators of compromise}
\end{acronym}

%% file: chapters/00-abstract.tex
\begin{abstract}
This paper presents a cost-effective and distributed flow monitoring platform for collecting unsampled IPFIX data exclusively using open-source tools, which is implemented at the University of Tübingen.
An overview of all tools is given and their use is explained.
\end{abstract}

%% file: content.tex
\section{Introduction}


Flow monitoring using IPFIX enables network operators to get insight into their network traffic.
This knowledge can be used for capacity planning or malicious traffic detection.
Many organizations that run flow monitoring rely on IPFIX flow exporters that are built into network devices.
To process every packet for flow generation, also called unsampled flow monitoring, devices need to be equipped with powerful CPUs.
However, these devices are expensive and can incur significant licensing fees.
A detailed introduction to flow monitoring is provided in Section \ref{ch:flowmon}.

This paper presents a cost-effective and distributed flow monitoring architecture for collecting unsampled IPFIX exclusively using open-source tools.
Such a flow monitoring architecture was implemented at the network at the University of Tübingen and is described in Section \ref{ch:case}.
It collects data from multiple strategic points within the network.
This enables the detection of flows inside the network, which would otherwise go unrecorded if monitoring is performed only at the network border.
By using only standard-compliant components for the flow monitoring platform, these components can be interchanged to meet specific requirements, such as the speed of the monitored traffic or the capabilities of the flow analysis pipeline. 


\section{A Primer on Flow Monitoring}
\label{ch:flowmon}

This section gives an overview of flow monitoring with IPFIX.
It presents different methods of flow generation and gives an introduction to flow collection.

\subsection{Overview}

Network traffic can be monitored efficiently by aggregating it into flows.
Each flow only contains metadata on the finished communication between two endpoints while discarding the actual payload.
In this way, the data that need to be processed is only a fraction of the original traffic volume.
Cisco originally started implementing NetFlow on their networking devices, which defined a protocol for transmitting flow information.
This was published as informational RFC 3954 \cite{rfc3954}.
The original NetFlow was designed to process every packet forwarded by a network device, i.e., unsampled NetFlow.
As the switching capacity surpassed the packet processing capabilities, sFlow (Sampled NetFlow) was introduced.
It works like NetFlow, but only takes every $n$-th packet into account, where $1/n$ is the so-called sampling rate.
RFC 7011 \cite{rfc7011} standardized the export of flows under the name IPFIX, based on the latest version of NetFlow.

Various papers discuss the usefulness of IPFIX for general traffic monitoring. 
Hofstede et al. \cite{6814316} provide an overview of all stages of a network monitoring setup using IPFIX.
They mention that traffic analysis using IPFIX is more efficient than packet-based monitoring, with a reduction in traffic of $1/2000$, and highlight its potential for various use cases, including intrusion detection, due to the extensibility of the IPFIX protocol.

\subsection{Flow Meters}

\begin{figure}[t]
    \centering
    \includesvg[width=\columnwidth]{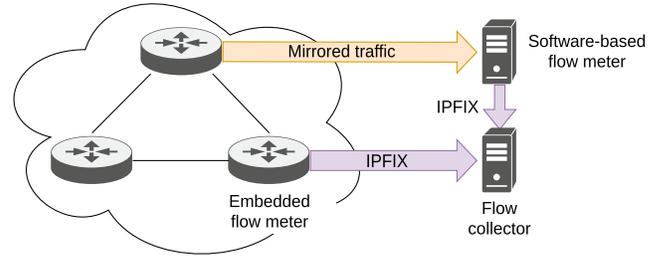}
    \caption{IPFIX collection methods.}
    \label{fig:ipfix_collection}
    \vspace{-0.5cm} 
\end{figure}

Flow meters generate flows from raw network traffic.
They can be categorized into two groups as shown in Figure \ref{fig:ipfix_collection}: Embedded flow meters and software-based flow meters.
Embedded flow meters are pre-installed on the switch or router itself.
They can be easily configured to export flows, but are often limited in performance and configuration options.
Furthermore, this feature is often marketed as a paid feature.
In contrast, software-based flow meters run on general-purpose server hardware and receive network traffic via a mirror port.
Mirror ports are generally supported by all enterprise-grade network devices, which allows network operators to choose any vendor.
Software-based flow meters also offer a wider range of configuration options and features, e.g., application labeling, deep packet inspection, and the choice of packet capture libraries.
To handle a higher traffic load, the hardware running the flow meter can be scaled accordingly.

\subsection{Flow Collectors}

Flow collectors receive flow data.
They can perform various functions in a flow monitoring platform, such as permanently store flow data, flow forwarding, and flow analysis.
The chosen storage format greatly affects the ability to analyze historic flow data.
Network operators are not limited to using a single flow collector.
As requirements for flow collection can vary, replicating IPFIX to multiple destinations allows for more flexible flow analysis.

\section{The SENSOR-Platform}
\label{ch:case}

This section describes the flow monitoring platform implemented at the University of Tübingen.
It discusses the network architecture and the components used.

\subsection{Overview}

\begin{figure}[t]
    \centering
    \includesvg[width=\columnwidth]{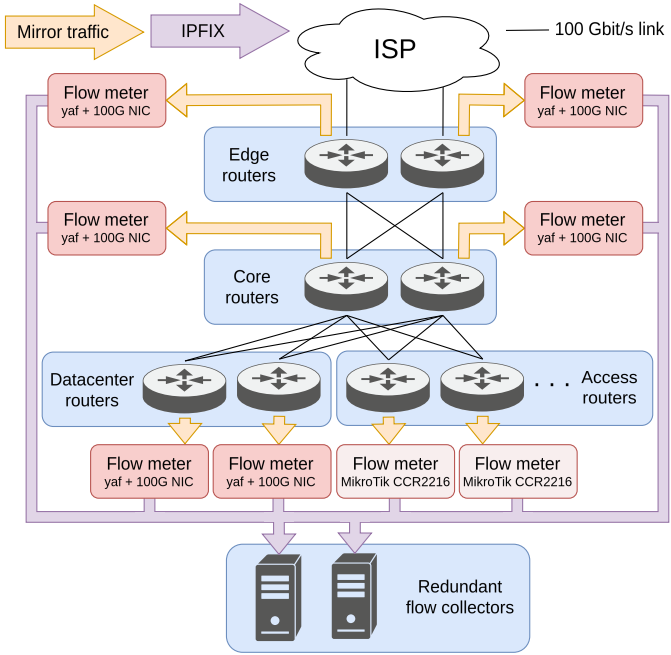}
    \caption{Flow monitoring platform at the University of Tübingen.}
    \label{fig:zdv_platform}
    \vspace{-0.5cm} 
\end{figure}

The university network shown in Figure \ref{fig:zdv_platform} comprises four types of routers: core, edge, datacenter, and access.
Instead of exporting flows via an embedded flow meter, dedicated flow meters are used at each of the routers.
Each meter receives a mirror of the forwarded traffic.
The generated IPFIX data is collected by a redundant setup of multiple flow collectors.
This increases resilience against hardware or software failures.


\subsection{Flow Meter Components}

As each router has a different traffic load, different flow metering methods are deployed.
For high-volume routers, such as those in the core, datacenter, and edge, a server equipped with an 100 Gbit/s \ac{nic} running the software-based flow meter \textit{yaf} is used.
For access routers with a lower traffic volume, the embedded flow meter of a MikroTik router is used.
This configuration is unusual because embedded flow meters typically export flows for forwarded network traffic.
However, the MikroTik router generates IPFIX using passively received mirror traffic.
Both methods of flow metering are described in the following.

\subsubsection{yaf}

\textit{yaf} \cite{yaf} is an open-source flow meter developed by the \ac{netsa} group at the CERT coordination center.
It can be configured to use the PF\_RING \cite{pfring} library for high-speed packet processing to handle incoming network traffic.
By implementing special optimization for a variety of \acp{nic}, it creates fewer internal copies while forwarding the packet to the application, thus increasing performance.
yaf also implements a plugin system which includes application labeling, \ac{dpi} capabilities, and identifying operating systems based on DHCP requests.
Further, plugins can be written to extend the functionality of yaf.

\subsubsection{MikroTik}
\label{sec:mikrotik}

MikroTik is a company that sells network devices with its own proprietary operating system, \textit{RouterOS}.
As their products do not offer the same range of features as those of major competitors, they are primarily used by individuals and small companies.
In terms of packet forwarding rate, however, these devices offer high performance compared to others in the same price range.
This makes them appealing to network operators for specific use cases.

In the context of flow metering, RouterOS is capable of exporting IPFIX with its \textit{traffic flow} feature.
Unfortunately, \textit{traffic flow} only processes packets, which are forwarded by the device, i.e., they traverse the network stack of RouterOS.
This makes it unsuitable for passive flow monitoring, where it receives a stream of mirrored traffic from another device without forwarding it.

\begin{figure}[t]
    \centering
    \includesvg[width=0.75\columnwidth]{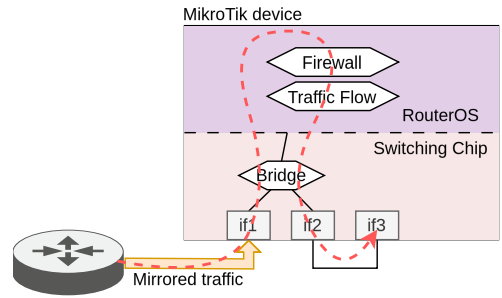}
    \caption{Physical setup for passive flow monitoring with MikroTik devices.}
    \label{fig:mikrotik_ipfixhack}
    \vspace{-0.5cm} 
\end{figure}

\begin{figure}[t]
\begin{lstlisting}[caption={RouterOS configuration for passive flow monitoring.},captionpos=b,label=lst:mikrotik_ipfixhack]
<@\textcolor{Fuchsia}{/interface bridge}@>
add arp=disabled name=mirror-br protocol-mode=none
<@\textcolor{Fuchsia}{/interface bridge port}@>
add bridge=mirror-br edge=no hw=no interface=if1 learn=no multicast-router=disabled point-to-point=no
add bridge=mirror-br edge=no hw=no interface=if2 learn=no multicast-router=disabled point-to-point=no
<@\textcolor{Fuchsia}{/interface bridge settings}@>
set use-ip-firewall=yes
\end{lstlisting}
\vspace{-0.7cm} 
\end{figure}

This restriction can be circumvented as follows.
First, as shown in Figure \ref{fig:mikrotik_ipfixhack}, we connect the mirror port of the monitored router with a free interface (\texttt{if1}) of the MikroTik device.
Then, a second link is established between any two free ports of the MikroTik router (\texttt{if2} and \texttt{if3}).
Next, a new bridge is configured on the device and \texttt{if1} and \texttt{if2} are attached to it.
The bridge is then configured to broadcast all incoming packets out of the remaining ports, regardless of any learned MAC addresses.
Until now, all configuration has been carried out on the switching chip, which is outside the scope of the traffic flow module.
Finally, the device is configured so that all packets passing through the bridge are subject to firewall checks.
This leads to all packets passing through the RouterOS network stack, making them subject to the \textit{traffic flow} module.
Once processed by \textit{traffic flow}, the packet is sent out via \texttt{if2}.
Finally, the packet is discarded at \texttt{if3} because this interface is not configured.
The complete RouterOS configuration described is shown in Listing \ref{lst:mikrotik_ipfixhack}.

This makes MikroTik devices a cost-effective option for flow metering.
The flow monitoring platform of the University of Tübingen uses a MikroTik CCR2216-1G-12XS-2XQ.
It was chosen because of its 100 Gbit/s port and its powerful 16-core CPU, which is important since the \textit{traffic flow} feature relies on the CPU.

\subsection{Flow Collector Components}

\begin{figure}[t]
    \centering
    \includesvg[width=0.9\columnwidth]{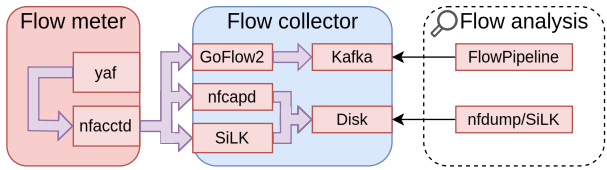}
    \caption{Flow collection components.}
    \label{fig:collection}
    \vspace{-0.5cm} 
\end{figure}

Each collector server consists of multiple flow-collecting processes as shown in Figure \ref{fig:collection}: nfcapd, SiLK flow collector, and GoFlow2.
Using multiple flow collectors allows flexible building of analysis pipelines.
In the following, all flow collectors components used in the distributed flow monitoring platform are briefly described.

\subsubsection{nfacctd}

nfacctd (NetFlow Accounting Deamon) \cite{pmacct} is an open-source program for processing NetFlow v9/IPFIX data.
In this setup, the \textit{tee plugin} can be used to replicate a single IPFIX stream to multiple other flow collectors.
This may be necessary since flow meters may not be able to send flow data to multiple destinations.
All exported flows by yaf are sent to the local nfacctd instance which then distributes them to the flow collectors.

\subsubsection{nfdump}

nfcapd (NetFlow Capture Daemon) is part of the nfdump project \cite{nfdump} which is a toolset for collecting and processing flow data.
The capture daemon receives IPFIX and stores the flow records in permanent storage.
Stored flows can then be analyzed with nfdump which supports filtering, aggregating, etc.
It supports filtering queries similar to those in tcpdump, making it easy for network operators to use.

\subsubsection{SiLK toolset}
SiLK (System for Internet-Level Knowledge) is a traffic analysis toolkit also developed by the \ac{netsa} group.
It includes its own flow collector \textit{rwflowpack} which stores incoming flows in an efficient binary format.
Other tools in the toolkit can be used to filter, aggregate, sort or match flows.
SiLK supports building complex operations by connecting multiple tools via Unix pipelines.
It also offers a Python library that allows direct interaction with stored flows.

\subsubsection{GoFlow2}

\textit{GoFlow2} \cite{goflow2} is a flow collector written in Golang.
Its predecessor \textit{GoFlow} was developed by Cloudflare.
It supports reading NetFlow, IPFIX, and sFlow and serializing them into a common output format.
The supported output formats are JSON and Protobuf, the latter of which is a binary serialization format developed by Google.
This allows for a simplified processing of flows further down the line since every flow has the same structure.
Conversion from NetFlow/IPFIX is necessary because the structure of flows in IPFIX messages is not fixed.
The IPFIX protocol requires templates to be sent periodically in order to define subsequent messages containing flow data.
The template structure depends on the flow meter used and how it is configured.
Once IPFIX has been converted into a defined output format, GoFlow2 writes it to a file or sends it to Apache Kafka, which is an event broker.
Each flow is emitted as an event, which can then be processed by other analysis components.

\subsubsection{flowpipeline}

Flowpipeline \cite{flowpipeline} is a GoFlow2-compatible flow processing toolkit.
It allows users to create pipelines for further flow processing.
Modules include data enrichment, anonymization, and metrics generation.
Flows can be read from Apache Kafka in GoFlow2 format.
This allows processing flow data in real time as Apache Kafka forwards events with minimal delay.

\section{Conclusion}
\label{ch:conclusion}

This paper presented how distributed and cost-effective unsampled flow monitoring is implemented at the University of Tübingen by only using vendor-agnostic open-source tools.
It also presented a configuration for MikroTik routers for generating IPFIX from an incoming stream of mirrored traffic.

While the flow meters used in this architecture can handle the traffic load of metered routers, the exact processing rates of these and many other flow meters remain unclear because they have not yet been tested.
To guarantee unsampled flow metering, the limits of the flow meters used should be well understood.
We will address this issue in future work, where we will measure the limits of various flow meters and their corresponding packet processing libraries.